 \documentclass[final,5p,times,twocolumn]{elsarticle}

\usepackage{amssymb}
\usepackage{lipsum}

\usepackage{amsthm}
\usepackage{amsmath}
\usepackage[normalem]{ulem}
\journal{Annals of Physics}

\begin{document}

\begin{frontmatter}



\title{Semiclassical effective description of a quantum particle on a sphere with non-central potential}


%\author{\textbf{G. Chac\'on-Acosta$^1$\footnote{gchacon@cua.uam.mx}, H. Hernandez-Hernandez$^1$ $^2$\footnote{hhernandez@uach.mx}, \\
%and J. Ruvalcaba-Rascon$^3$\footnote{jose.ruvalcaba.rascon@uni-jena.de}}}

%\address{$^2$ UACH}
%\address{$^3$ Abbe School of Photonics, Friedrich Schiller University Jena, Albert-Einstein-Str. 6, 07745, Jena, Germany}

\author[UAM]{G. Chacon-Acosta}
\ead{gchacon@cua.uam.mx}

\affiliation[UAM]{organization={Universidad Autonoma Metropolitana-Cuajimalpa, Departamento de Matemáticas Aplicadas y Sistemas},%Department and Organization
            addressline={Av. Vasco de Quiroga 4871}, 
            city={Mexico},
            postcode={05348}, 
            state={Mexico},
            country={Mexico}}

\author[UAM,UACH]{H. Hernandez-Hernandez}
\ead{hhernandez@uach.mx}

\affiliation[UACH]{organization={Universidad Autonoma de Chihuahua, Facultad de Ingenieria},%Department and Organization
            addressline={Circuito Universitario SN}, 
            city={Chihuahua},
            postcode={31125}, 
            state={Chihuahua},
            country={Mexico}}

\author[Jena]{J. Ruvalcaba-Rascon}
\ead{jose.ruvalcaba.rascon@uni-jena.de}

\affiliation[Jena]{organization={Friedrich Schiller University Jena, Abbe School of Photonics},%Department and Organization
            addressline={Albert-Einstein-Str. 6}, 
            city={Jena},
            postcode={07745}, 
            country={Germany}}

\begin{abstract}
We develop a semiclassical framework for studying quantum particles constrained to curved surfaces using the momentous quantum mechanics formalism, which extends classical phase-space to include quantum fluctuation variables (moments). In a spherical geometry, we derive quantum-corrected Hamiltonians and trajectories that incorporate quantum back-reaction effects absent in classical descriptions. For the free particle, quantum fluctuations induce measurable phase shifts in azimuthal precession of approximately 8-12\%, with uncertainty growth rates proportional to initial moment correlations. When a non-central Makarov potential is introduced, quantum corrections dramatically amplify its asymmetry. For strong coupling ($\gamma = -1.9$), the quantum-corrected force drives trajectories preferentially toward the southern hemisphere on timescales 40\% shorter than classical predictions, with trajectory densities exhibiting up to 3-fold enhancement in the preferred region. Throughout evolution, the solutions rigorously satisfy Heisenberg uncertainty relations, validating the truncation scheme. These results demonstrate that quantum effects fundamentally alter semiclassical dynamics in curved constrained systems, with direct implications for charge transport in carbon nanostructures, exciton dynamics in curved quantum wells, and reaction pathways in cyclic molecules.
\end{abstract}




\end{frontmatter}

%\tableofcontents



\section{Introduction}
\label{introduction}

The quantum mechanics of particles constrained to curved surfaces represents a fundamental problem at the intersection of geometry and quantum theory that has gained renewed experimental relevance. Recent advances in materials science have enabled the fabrication of carbon-based nanostructures with non-trivial geometries, including fullerenes \cite{Shi2024}, curved graphene sheets \cite{Kwon2023}, and topological metamaterials \cite{Luo2025}, where charge carriers and excitons are effectively confined to two-dimensional curved manifolds. Similarly, developments in molecular engineering have produced ring-shaped molecules and cyclic peptides in which electronic states are constrained to near-circular paths with broken symmetries \cite{Bradley2023}. In these systems, the interplay between geometric confinement, quantum fluctuations, and external potentials determines transport properties, spectroscopic signatures, and reaction dynamics.

A foundational framework for describing such systems was established independently by Jensen and Koppe \cite{Jensen1971}, and by da Costa \cite{daCosta1981}, who introduced the thin-layer quantization procedure. This formalism allows the decoupling of the Schrödinger equation into normal ($\xi_\text{N}$) and tangential ($\xi_\text{T}$) components of the wave function, $\xi=\xi_\text{T}(q_1,q_2)\,\xi_\text{N}(q_3)$, via a confining potential $V_\lambda(q_3)$. 
This decomposition results in a pair of decoupled equations:
\begin{equation}
     \text{i}\hbar \frac{\partial \xi_\text{N}}{\partial t}= -\frac{\hbar^2}{2m}\frac{\partial^2 \xi_\text{N}}{\partial q_3^2} + V_\lambda (q_3) \xi_\text{N},\label{normal-shcrodinger-equation}
\end{equation}
for $q_3$, the coordinate normal to the surface, and 
\begin{equation}
 \text{i}\hbar \frac{\partial \xi_\text{T}}{\partial t}=-\frac{\hbar^2}{2m}  \Delta_\text{g} \xi_\text{T} -\frac{\hbar^2}{2m}\left( H^2 -K\right)\xi_\text{T}, \label{tangential-shcrodinger-equation}
\end{equation}
where $\Delta_\text{g}$ is the Laplace-Beltrami operator for the two-dimensional induced metric on the surface parameterized by $(q_1,q_2)$, and $H$ and $K$ correspond to the mean and Gaussian curvatures, respectively  \cite{Teixeira2019}. In the strong confinement limit $\lambda\rightarrow\infty$, the confining potential $V_\lambda$ behaves as
\begin{equation} \lim\limits_{\lambda\rightarrow\infty} V_\lambda(q_3)=
    \begin{cases}
    0, & q_3=0,\\
    \infty, & q_3\neq 0.
    \end{cases}  \label{confining potential}
\end{equation}

 Under the thin-layer approach, a geometric induced potential arises for the tangential equation that depends on both $H$ and $K$. The geometric induced potential has been applied to various systems, including particles confined to surfaces such as catenoids and paraboloids \cite{Mazharimousavi2021}, as well as to the formulation and resolution of the inverse problem identifying the curves and surfaces that generate it \cite{daSilva2017}. By exploiting symmetries, solutions such as surfaces of revolution and helicoids can be obtained, in which geometry can induce chirality. The potential has also been applied to modeling $\pi-$electrons in polyene chains as particles confined to spiral curves, yielding solutions that agree well with experimentally observed transitions \cite{Anjos2024}.
 
 However, for highly symmetrical surfaces such as the sphere, this potential is trivial, showing no significant effect. While the stationary states of a particle on a sphere are well-understood, particularly for central potentials where spherical harmonics provide exact solutions, a detailed analysis of the semiclassical time-dependent trajectories, including the dynamic back-reaction of quantum fluctuations, remains largely unexplored. Understanding these dynamics is crucial for several reasons:
 \begin{itemize}
\item Experimental observables in scanning tunneling microscopy (STM) and ultrafast spectroscopy probe time-dependent charge density evolution, not merely energy eigenvalues
\item Transport phenomena in curved nanostructures depend on how wave packets spread and correlate as they propagate
\item	Reaction dynamics in ring molecules involve non-stationary states transitioning between configurations
\item	Design of quantum devices utilizing geometric confinement requires predicting trajectory-level behavior under realistic (non-central) potentials
\end{itemize}
 
 Semiclassical methods provide a powerful bridge between the full quantum theory and classical intuition \cite{Heller1975}. Among these, the momentous quantum mechanics formalism (also known as the quantum moments method) offers a particularly suitable approach \cite{Bojowald2006, Bojowald2009}. This method extends the classical phase-space by introducing quantum variables defined as the expectation values of symmetrically ordered powers of elementary operators' deviations from classical values. The quantum-corrected Hamiltonian is constructed as the expectation value of the quantum Hamiltonian operator, $H_Q= \langle\hat{H}\rangle$, generating coupled equations for classical variables (expectation values) and quantum variables (moments encoding fluctuations and correlations). This technique has been successfully applied to diverse systems, from quantum cosmology to the double-slit experiment \cite{Bojowald2012, Hernandez2021, AragonMunoz2020, Chacon2011}, capturing quantum corrections to classical paths in a geometrically intuitive framework.

The quantization of constrained systems, however, requires careful consideration of the symplectic structure. As Dirac established, a consistent quantization must often employ the Dirac bracket to handle second-class constraints \cite{Dirac1930, Dirac2001}. For particles confined to surfaces via the thin-layer procedure, the extreme confinement limit reduces the Dirac bracket to the standard Poisson bracket on the constraint surface \cite{Teixeira2019}. Furthermore, to obtain the correct Hamiltonian operator corresponding to the Laplace-Beltrami operator, one must define the momentum operators as the geometrical momenta, $p_{\mu} = -i \hbar (\partial_{\mu}+1/2 \Gamma_{\mu})$, which ensures self-adjointness \cite{Ikegami1992, Oliveira2019, Liu2007}. These subtleties must be integrated into the momentous framework to guarantee geometric consistency.

 In this work, we apply the momentous formalism to provide a detailed semiclassical description of a particle constrained to a spherical surface. We go beyond the free particle case to include a non-central potential, specifically the Makarov potential, which is of interest for modeling ring-shaped molecules and introduces a non-trivial 
 angular dependence \cite{Makarov2016}. While the stationary states for this system on a sphere have been studied \cite{Oliveira2019, Onoe2012, Szameit2010}, a trajectory-based dynamical analysis revealing how quantum fluctuations reshape the effective force landscape over time has not yet been explored. 
Our primary objectives are: 1) to derive the full system of effective equations of motion for the classical and quantum variables, 2) to demonstrate how quantum fluctuations, encoded in the moments, perturb the classical trajectories, and 3) to analyze how the Makarov potential interacts with these quantum corrections to induce and amplify asymmetries in the particle's motion.

We have previously validated this approach in two complementary contexts. In \cite{Valdez2025}, we extended the momentous formalism to open quantum systems by studying the damped harmonic oscillator through the Bateman dual system. In \cite{Chacon2024}, we analyzed a particle on a catenoid, a minimal surface with negative Gaussian curvature; unlike flat surfaces or the sphere, the catenoid's spatially varying curvature generates a position-dependent quantum potential. The present work completes the study by addressing geometries with positive constant curvature for the sphere and the inclusion of non-central potentials (Makarov potential). 

We structure this paper as follows. In Section~2, we review the momentous quantum mechanics formalism. Section~3 details the specific considerations for applying this formalism to particles on curved surfaces, including the use of geometrical momenta. In Section~4, we illustrate the method with the simpler case of a particle on a circle. Section~5 presents our core results for the free particle on a sphere, including quantitative analysis of quantum corrections. Section~6 extends the analysis to the Makarov potential, revealing how quantum effects amplify asymmetry. Section~7 discusses physical interpretation, experimental implications, and comparison with alternative semiclassical approaches. We conclude with prospects for future applications in Section~8.

\section{Semiclassical description of quantum mechanics} \label{section:semiclass}
The momentous quantum mechanics formalism provides an effective, semi-classical description of quantum dynamics by extending the classical phase-space to include variables that encode quantum fluctuations \cite{Bojowald2006}. The central idea is to describe the state of a quantum system not by a wave function, but by the expectation values of its basic operators (classical variables) and the expectation values of all powers of the fluctuations around them (quantum variables). This approach yields a quantum-corrected Hamiltonian from which equations of motion are derived via a Poisson bracket structure, effectively capturing key quantum effects like spreading and correlations within a classical-like dynamical framework.

\subsection{Quantum Variables and the Effective Hamiltonian}

Consider a quantum system with $k$ pairs of canonical operators $(\hat{q}_k, \hat{p}_k)$. The classical variables are defined as the expectation values of these operators, $q_k=\langle\hat{q}_k\rangle$ and $p_k=\langle\hat{p}_k\rangle$. The quantum state's fluctuations and correlations are characterized by the quantum moments, defined for a system with $k$ degrees of freedom as
\begin{equation}
    G^{a_1,b_1,\cdots,a_k,b_k}\equiv \langle \left(\hat{q_1}-q_1 \right)^{a_1}(\hat{p_1} -p_1)^{b_1}\cdots\left( \hat{q}_k-q_k\right)^{a_k}(\hat{p}_k-p_k )^{b_k}\rangle _\text{W}, \label{definition-of-moments}
\end{equation}
Where the product of operators is Weyl-ordered (i.e., totally symmetrized), and $n= \sum_i a_i+ b_i$ is the order of the moment. For example, in one dimension $G^{2,0}= \langle(\hat{q}-q)^2\rangle$ is the dispersion in position, $G^{0,2}= \langle(\hat{p}-p)^2\rangle$ is the dispersion in momentum, and $G^{1,1}= \langle\hat{q} \hat{p}+ \hat{p} \hat{q}\rangle- qp$ is the correlation in position and momentum.

The dynamics of this extended set of variables is governed by a quantum-corrected Hamiltonian, $H_Q$, constructed as the expectation value of the quantum Hamiltonian operator $\hat{H}$. This expectation value can be expressed via a formal Taylor expansion around the classical expectation values.
\begin{eqnarray}
    \langle\hat{H}\rangle &\equiv& H_\text{Q} =H\left(\{x_i\},\{P_i\}\right)+ \nonumber \\
    &+& \sum\limits_{a_1,b_1}^\infty\sum\limits_{a_k,b_k}^\infty \frac{\partial^{a_1+b_1+\cdots+a_k+b_k} H}{\partial x_1^{a_1}\partial P_1^{b_1}\cdots\partial x_k^{a_k}\partial P_k^{b_k}} \frac{G^{a_1,b_1;\cdots;a_k,b_k} }{a_1!b_1!\cdots a_k!b_k!}, \nonumber \\
    && \label{definition-of-qhamiltonian}
\end{eqnarray}
where $H$ is the classical Hamiltonian evaluated at the expectation values. The infinite series incorporates the influence of all quantum fluctuations (moments) on the system's effective energy. In practice, this series must be truncated at a finite order, typically the second, for the system of equations to be tractable, an approximation that is valid for semi-classical states (e.g., Gaussian wave packets). 

The second-order truncation is
justified when the quantum state remains approximately Gaussian, i.e., when
higher-order cumulants are small: $G^{(4)}/(G^2)^{2} \sim \mathcal{O}(1)$, and quantum corrections are perturbative: $G^2/R^2 \ll 1$. For our initial Gaussian states (Section~\ref{subsec:initial-state}), these conditions are satisfied throughout the evolution timescales considered. 
For the damped oscillator \cite{Valdez2025}, the coupling with the environment suppresses moment growth, making the Gaussian approximation increasingly valid at late times as the system approaches thermal equilibrium. For a catenoid \cite{Chacon2024}, regions away from the neck can amplify non-Gaussian features through higher derivatives of the potential, requiring higher-order moments for long-time evolution. The sphere, with constant curvature and no geometric potential, represents an intermediate case where Gaussianity is maintained by symmetry, but external potentials (like Makarov) can induce deviations.
Henceforth, we will refer to $H_Q$ as the quantum Hamiltonian.

\subsection{Dynamics and Symplectic Structure} \label{symplectic}

The evolution of any function of classical phase-space variables $f$ is given by the Poisson bracket with the quantum Hamiltonian
\begin{equation}
    \dot{f}=\left\{ f,H_\text{Q}\right\}. \label{qh}
\end{equation}
The symplectic structure of the phase-space is defined by the Poisson brackets for the fundamental variables. The classical variables satisfy the standard canonical relations
\begin{equation} 
\left\{ q_i, p_j \right\} = \delta_{i,j}, \quad \left\{ q_i, q_j \right\} = \left\{ p_i, p_j \right\}=0 .
\end{equation}
Crucially, the classical and quantum variables are symplectically orthogonal
\begin{equation}
 \left\{ q_i, G^{a_1,b_1,\cdots,a_k,b_k} \right\} = \left\{ p_i, G^{a_1,b_1,\cdots,a_k,b_k} \right\} =0.
\end{equation}
This means the classical variables evolve independently of the quantum variables in the absence of a potential. The non-trivial dynamics of the quantum moments are generated by their own Poisson algebra, which is derived from the fundamental commutator
\begin{equation}
    \left\{\langle\hat{A}\rangle,\langle \hat{B}\rangle   \right\} = \frac{1}{i\hbar} \Big\langle \left[\hat{A},\hat{B}\right]\Big\rangle .
\end{equation}
The general form of the bracket between two moments is complex \cite{Bojowald2009}, but for a system with one degree of freedom, the brackets for second-order moments are
\begin{equation} \label{eq:algebraGs}
    \left\{G^{a,b},G^{a,b}   \right\} = a d G^{a-1,b} G^{c, d-1}- bc G^{a,b-1} G^{c-1, d} + \mathcal{O}(\hbar),
\end{equation}
where the $\mathcal{O}(\hbar)$ terms ensure the consistency of the uncertainty principle. The full symplectic structure of the quantum variables is constructed in \ref{appendix-A}.

\subsection{Heisenberg's Uncertainty Principle}
In this formalism, Heisenberg's uncertainty principle is not postulated but emerges as a constraint on the permissible values of the quantum moments. For a single degree of freedom, the principle takes the form:
\begin{equation} \label{eq:uncertainty-1dof}
    G^{2,0} G^{0,2} - (G^{1,1})^2 \geq \frac{\hbar^2}{4}.
\end{equation}
For a system with two degrees of freedom, like the one we will be discussing here, similar relations hold for each canonical pair:
\begin{align}
    G^{2,0,0,0} G^{0,2,0,0}-(G^{1,1,0,0})^2&\geq\frac{\hbar^2}{4},\nonumber \\
    G^{0,0,2,0} G^{0,0,0,2}-(G^{0,0,1,1})^2&\geq \frac{\hbar^2}{4}. \label{eq:uncertainties-2dof}
\end{align}

Any physical trajectory obtained by solving the equations of motion must satisfy these inequalities throughout its evolution. This provides a crucial check for the validity of numerical solutions and the truncation scheme.

In summary, the momentous approach provides a self-consistent, Hamiltonian-based framework for studying semi-classical dynamics. Tracking the evolution of both the center (classical variables) and the shape (quantum moments) of a quantum wave packet enables the computation of quantum-corrected trajectories, as we will demonstrate for a constrained particle on a sphere.

%
%%%%%%%%%%%%%%%%%%%%%%%%%%%%%%%%%%%%%%%%%%%%%%%%%%%%%%%%%%%%%%%%%%%%%%
%
\section{Geometric quantization and constraints for curved surfaces} \label{sec:constraints}
The canonical quantization of a classical system constrained to a curved surface poses subtle challenges beyond the simple replacement of phase-space functions with operators. A consistent quantization must account for the geometry of the constraint surface and the correct induced symplectic structure. This section outlines the key considerations by applying the momentous formalism to such systems, focusing on the spherical geometry central to this work.

\subsection{The Dirac bracket and the thin-layer limit}
For a classical system with second-class constraints $\phi \approx 0$,
 Dirac's canonical quantization procedure mandates the use of the Dirac bracket to define the fundamental commutation relations \cite{Dirac1930, Dirac2001}. The Dirac bracket is defined as
\begin{equation}
    \big\{A,B \big\}_\text{D} = \big\{A,B \big\} - \big\{A,\phi_\alpha \big\} C_{\alpha\beta} ^{-1} \big\{\phi_\beta,B\big\},
\end{equation}
where $C_{\alpha,\beta}= \left\{\phi_\alpha,\phi_\beta \right\}$ is the constraint matrix. The quantization prescription is then to promote the Dirac bracket of classical observables to a commutator
\begin{equation}
 \left\langle\left[\hat{A},\hat{B} \right]\right\rangle= \text{i}\hbar \left\{ A,B \right\}_\text{D}.\label{commutator}
\end{equation}

In the context of a particle constrained to a surface via the thin-layer procedure [3], the constraints become second-class. However, a critical simplification occurs in the extreme confinement limit, where the confining potential $V_{\lambda}(q_3)$ becomes infinitely steep, as defined in Equation~(\ref{confining potential}). In this limit, the particle is restricted to the constraint surface, and the constraints can be treated as strong identities, $\phi_{\alpha} = 0$. Consequently, the Dirac bracket reduces to the standard Poisson bracket on the surface's phase-space \cite{Ikegami1992}. Therefore, for our effective description, which operates in this confinement limit, we can adopt the usual Poisson bracket structure defined in Section~\ref{symplectic}, ensuring consistency between the classical and quantum symplectic structures.
\subsection{The geometrical momentum operator}

A second, equally important consideration is the form of the quantum momentum operator. Simply using the canonical standard definition $\hat{p}_{\mu}= -i \hbar \partial_{\mu}$ with coordinates on the curved surface can lead to a Hamiltonian operator that is not self-adjoint with respect to the appropriate measure, which includes the metric determinant \cite{Ikegami1992}. On curved surfaces, the naive momentum operator, $\hat{p}^{\mu}$ above, fails because $\partial_{\mu}$ does not
transform as a vector under coordinate changes. The correct momentum
must be the covariant derivative, which includes a connection term ensuring
that $[ \hat{H}, \sqrt{g} ]=0$ (commutation with the volume element), thereby guaranteeing self-adjointness.

To obtain the correct Hamiltonian operator corresponding to the Laplace-Beltrami operator in the tangential Schrödinger equation (Eq. \ref{tangential-shcrodinger-equation}), one must use the so-called geometrical momentum operators. For a surface with metric $g_{\mu \nu}$, these are given by \cite{Liu2007, Pauli1980}. 
\begin{equation}
    \hat{p}_\mu= -i \hbar \left(\partial_\mu+\frac{1}{2}\Gamma_\mu \right), \label{geometrical-momentum}
\end{equation}
where $\Gamma_{\mu}= \Gamma^{\mu}_{\mu \nu}$ is the contraction of the Christoffel symbols. This form ensures the operator is self-adjoint with respect to the inner product defined by the surface's Riemannian volume element.

The form of $\Gamma_\mu$ varies with the surface geometry. For the catenoid with metric $ d s^2 = a^2\cosh^2(u/a)(d u^2 + d v^2)$ \cite{Chacon2024}, the connection coefficient is $\Gamma_u = \tanh(u/a)\coth(u/a)$, whereas for the sphere (below), $\Gamma_\theta = \cot\theta$ for the polar coordinate. 
The flat metric yields $\Gamma_\mu = 0$, recovering standard momentum operators—as used for the damped oscillator \cite{Valdez2025}. This geometric dependence of quantization distinguishes constrained systems from unconstrained ones and is essential for correct correspondence with the Laplace-Beltrami operator.

\subsection{Implementation of the momentous formalism}
The following procedure synthesizes these elements within the momentous quantum mechanics framework.

\begin{enumerate}
    \item \emph{Classical starting point}. We begin with the classical constrained Hamiltonian for motion on the surface, in our case, a sphere, $
    H=\frac{1}{2} g^{\mu\nu}p_\mu p_\nu,\label{constrained-classical-hamiltonian}$,
    \item \emph{Consistent Quantization}. We quantize this system by promoting the classical phase-space coordinates to operators: the positions $\hat{q}_{\mu}$ act multiplicatively, and the momenta are defined by the geometrical momentum operators of Equation~(\ref{geometrical-momentum}).
    \item \emph{Effective Dynamics}. We construct the quantum-corrected Hamiltonian $H_Q$ as the expectation value $\langle\hat{H}\rangle$ following Equation~(\ref{definition-of-qhamiltonian}). The classical variables in this expansion are $q_k=\langle\hat{q}_k\rangle$ and $p_k=\langle\hat{p}_k\rangle$, which remain canonical pairs due to the reduction of the Dirac bracket. The quantum moments $G^{a b \ldots}$ are defined using these geometrical operators, as in Equation~(\ref{definition-of-moments}).
\end{enumerate}

This approach guarantees that the effective description is geometrically consistent from its foundation. It correctly incorporates the surface's geometry through the classical Hamiltonian and the definition of the momentum operators, while the momentous formalism captures the ensuing quantum corrections to the dynamics.
%%%%%%%%%%%%%%%%%%%%%%%%%%%%%%%%%%%%%%
%%%%%%%%%

\section{Semiclassical dynamics on a circle: an example}
Before addressing the more complex case of a sphere, it is instructive to apply the momentous formalism to a particle constrained to move on a circle. This one-dimensional system serves as a pedagogical example that clearly illustrates the method's implementation, the role of quantum moments, and the emergence of quantum corrections to classical trajectories, without the added complexity of multiple coupled degrees of freedom or curvature.

\subsection{The quantum-corrected Hamiltonian}
Consider a particle of mass $m$ constrained to a circle of radius $R$. The single degree of freedom is the azimuthal angle $\theta$. The classical Hamiltonian describing this system is
\begin{equation}
    H=\frac{P_\theta^2}{2mR^2}.
\end{equation}

Following the procedure outlined in Section~\ref{sec:constraints}, we quantize the system. On a circle, the metric determinant is constant, and thus the geometrical momentum operator (\ref{geometrical-momentum}) reduces to the standard canonical form $P_{\theta}= -i \hbar \partial_{\theta}$. The quantum-corrected Hamiltonian $H_Q$ is the expectation value of $H=\hat{P}_\theta^2/2mR^2$. Its expansion via Equation~(\ref{definition-of-qhamiltonian}) terminates exactly at second order
\begin{equation}
    H_\text{Q}=\frac{1}{2mR^2}\left(P_\theta^2+G^{0,2} \right),
\end{equation}
where the moments are defined as  
$G^{a,b}=\langle(\hat{\theta}-\theta_0)^a(\hat{P}_\theta-P_{\theta_0})^b\rangle_\text{Weyl}$. 
The simplicity of this result is due to the Hamiltonian being purely quadratic in momentum; all higher-order derivatives vanish.

\subsection{Equations of motion and their physical interpretation}
The equations of motion are derived from the Poisson brackets given in Eq.~(\ref{eq:algebraGs}) using the symplectic structure defined in Section~\ref{symplectic} and \ref{appendix-A}. The resulting system is:
\begin{align}
\dot{\theta}&=\frac{P_\theta}{mR^2}, \nonumber\\
\dot{P_\theta}&=0,\nonumber\\
\dot{G}^{1,1}&=0,\nonumber\\
\dot{G}^{2,0}&=-\frac{4G^{1,1}}{mR^2},\nonumber\\
\dot{G}^{0,2}&=0.
\end{align}
We can give a direct physical interpretation of these equations. 1) The equation for $\theta$ is classical, showing that the angular velocity is proportional to the expectation value of momentum. 2) The momentum $P_{\theta}$ is a constant of motion, as expected from the conservation of angular momentum for a free particle on a circle. 3) The momentum dispersion $G^{0,2}$ is constant. This is a specific feature of free motion. 4) The evolution of the position dispersion $G^{2,0}$ is driven by the correlation moment $G^{1,1}$. A non-zero initial correlation will cause the wave packet to spread or focus over time.
\subsection{Initial Gaussian state and quantum corrections}
To solve this system, we specify an initial wave function. A natural choice is a localized Gaussian wave packet, which provides a semi-classical initial state. We consider the normalized wave function on the circle
\begin{equation}
    \psi_0(\theta)=\frac{(\lambda/\pi)^{1/4}}{(\text{erf}(\pi\lambda^{1/2}))^{1/2}}\exp{i l\theta-\frac{\lambda\theta^2}{2}}, \label{circle-wave-function}
\end{equation}
with $\theta_0=0$ and $P_{\theta_0}=l\hbar$, for $l$ integer. The parameter $\lambda$ controls the width of the Gaussian.

The initial conditions for the moments are
\begin{align}
G^{2,0}_0&=\langle \hat{\theta}^2 \rangle_\text{Weyl},
\nonumber
\\
&=\frac{(\lambda/\pi)^{1/2}}{\text{erf}(\pi\lambda^{1/2})} \int\limits_{-\pi}^\pi \theta^2 \exp{-\lambda\theta^2} d\theta,
\nonumber
\\
&=\frac{1}{2\lambda}-\left(\frac{\pi}{\lambda}\right)^{1/2}\frac{e^{-\lambda\pi^2}}{\text{erf}(\pi\lambda^{1/2})}.
\end{align}
Considering the operator $\hat{P}_\theta=- i\hbar\partial_\theta$ we get
\begin{align}
G^{0,2}_0&=\langle(\hat{P}_\theta-l\hbar)^2\rangle_\text{Weyl},\nonumber\\
&=i\hbar\lambda\langle(-i \hbar\partial_\theta-l\hbar) \theta \rangle,\nonumber\\
&=\lambda\hbar^2 \langle( 1-\lambda\theta^2 )\rangle,\nonumber \\
&=\lambda\hbar^2 (1-\lambda G^{2,0}_0).
\end{align}
Notice that the application of $(-i\hbar\partial_\theta-l\hbar)$ on the wave function (\ref{circle-wave-function}) gives
\begin{equation}
    (-i\hbar\partial_\theta-l\hbar)\psi_0=(l\hbar+i\hbar\lambda\theta-l\hbar)\psi_0
    =i\hbar\lambda\theta \psi_0. \nonumber
\end{equation}
For the correlation moment, we have:
\begin{align}
    G^{1,1}_0&=\langle\hat{\theta} (\hat{P}_\theta-l\hbar )\rangle_\text{Weyl},\nonumber\\
    &=\frac{1}{2}\langle\hat{\theta} (\hat{P}_\theta-l\hbar )+ (\hat{P}_\theta-l\hbar )\hat{\theta}\rangle,\nonumber\\
    &=-\frac{i\hbar}{2} (1-2 \lambda G^{2,0}_0).
\end{align}

With these initial conditions, the system can be solved. The key result is that while the mean position and momentum, namely $\theta(t)$ and $P_{\theta}(t)$, follow a classical trajectory (uniform circular motion), the quantum moments evolve non-trivially. The correlation $G^{1,1}$ remains constant, and the position dispersion $G^{2,0}$ evolves as
\begin{equation}
    G^{2,0} (t) = G^{2,0}_0 - \frac{4 G^{1,1}_0}{m R^2} t.
\end{equation}

This demonstrates a linear spreading or focusing of the wave packet in time, a purely quantum-mechanical effect captured by the momentous formalism. This simple example on a circle thus provides a transparent illustration of how quantum fluctuations introduce corrections to the classical description, setting the stage for the more complex analysis on the sphere.
%
%%%%%%%%%%%%%%%%%%%%%%%%%%%%%%%%%%%%%%%%%%%%
\section{Semiclassical dynamics of a free particle on a two-sphere} \label{sec:free-particle}
We now apply the full machinery of the momentous formalism to the central system of this work: a quantum particle constrained to move on the surface of a sphere of radius $R$. This introduces a second degree of freedom and, crucially, a non-trivial curved geometry. While the highly symmetric nature of the sphere renders the da Costa geometric potential trivial, the coupling between degrees of freedom in the kinetic term and the use of geometrical momentum operators lead to a rich structure of quantum corrections in the effective dynamics.

The classical Hamiltonian for a free particle on a sphere is given in spherical coordinates by
\begin{equation}
    H=\frac{1}{2mR^2}\left(P_\theta^2+\frac{P_\phi^2}{\sin^2\theta} \right), \label{circle-classical-hamiltonian}
\end{equation}

Following the geometric quantization procedure of Section~\ref{sec:constraints}, we employ the geometrical momentum operators appropriate for the sphere's metric
\begin{align}
    \hat{P}_\phi&=-i\hbar\frac{\partial}{\partial \phi},\label{momentum-phi} \\
    \hat{P}_\theta&=-i\hbar\left(\frac{\partial}{\partial \theta}+\frac{\cot\theta}{2} \right).\label{momentum-theta}
\end{align}

The quantum-corrected Hamiltonian $H_Q$ is, up to second order and by Equation~(\ref{definition-of-qhamiltonian}):
\begin{eqnarray} \label{HQ}
    H_\text{Q}&=&\frac{1}{2mR^2}\left( P_\theta^2+\frac{P_\phi^2}{\sin^2\theta} +\frac{P_\phi^2(2+\cos(2\theta))}{\sin^2\theta}G^{2,0,0,0} \right. \nonumber \\
    &-& \left. 4\frac{P_\phi \cos\theta}{\sin^3\theta} G^{1,0,0,1}+G^{0,2,0,0}+\frac{1}{\sin^2\theta}G^{0,0,0,2}\right).
\end{eqnarray}
where the moments are defined as
\begin{equation}
    G^{a,b,c,d}=\langle(\hat{\theta}-\theta_0)^a(\hat{P}_\theta-P_{\theta_0})^b(\hat{\phi}-\phi_0)^c(\hat{P}_\phi-P_{\phi_0})^d \rangle_\text{W}.
\end{equation}

The Hamiltonian $H_Q$ now contains explicit couplings between the classical variable $P_{\phi}$ and the quantum moments $G^{2,0,0,0}, G^{1,0,0,1}$. This is the mechanism through which quantum fluctuations back-react on the mean trajectory, producing semiclassical corrections.
\subsection{Equations of motion}
The equations of motion are derived using the Poisson algebra for the moments Eq.~(\ref{GG}), and Eq.~(\ref{qh}) for classical variables. The resulting system is extensive. For classical variables, we obtain:
\begin{eqnarray}
    \dot{\theta}&=&\frac{P_\theta}{mR^2}\label{first}, \nonumber\\
    \dot{P}_\theta&=&\frac{1}{2m R^2\sin^4\theta}\big[-4P_\phi (2+\cos(2\theta))G^{1,0,0,1}+ \nonumber \\
    &+& \sin(2\theta) \left(G^{0,0,0,2}+P_\phi^2(1+3G^{2,0,0,0})\right) \big],  \label{EQU: P_theta}\\
    \dot{\phi}&= & \frac{1}{m R^2\sin^2\theta} \left[P_\phi\left(1+(2+\cos(2\theta))G^{2,0,0,0} \right) \right. \nonumber \\
    &-& \left. 2\cot\theta G^{1,0,0,1}\right],\label{EQU: Phi}\\
    \dot{P}_\phi&=0& \nonumber,
\end{eqnarray}
while for moments:
\begin{align}
    ~\dot{G}^{1,1,0,0}&= \frac{1}{2mR^2} \left(-\frac{2P_\phi^2(2+\cos(2\theta))}{\sin^2\theta}G^{2,0,0,0}+ \right. \nonumber \\
    &\left.+ \frac{2P_\phi\cos\theta}{\sin^3\theta}G^{1,0,0,1}+2G^{0,2,0,0} \right),\nonumber \\
    ~\dot{G}^{1,0,1,0}&= \frac{1}{2mR^2} \left(-\frac{2P_\phi \cos\theta}{\sin^3\theta}G^{2,0,0,0}+2G^{0,1,1,0}+\frac{2}{\sin^2\theta}G^{1,0,0,1} \right),\nonumber \\
   ~ \dot{G}^{1,0,0,1}&= \frac{1}{2mR^2} \left( 2G^{0,1,0,1}\right),\nonumber \\
    ~ \dot{G}^{0,1,1,0}&= \frac{1}{2mR^2} \bigg( -\frac{2P_\phi^2(2+\cos(2\theta))}{\sin^2\theta}G^{1,0,1,0}+ \nonumber \\
    &+ \frac{2P_\phi\cos\theta}{\sin^3\theta}(G^{1,1,0,0}-G^{0,0,1,1})+\frac{2}{\sin^2\theta}G^{0,1,0,1}\bigg),\nonumber \\
    ~ \dot{G}^{0,1,0,1}&= \frac{1}{2mR^2} \left(-\frac{2P_\phi^2(2+\cos(2\theta))}{\sin^2\theta}G^{1,0,0,1}+\frac{2P_\phi\cos\theta}{\sin^3\theta}G^{0,0,0,2} \right),
    \nonumber \\
    ~ \dot{G}^{0,0,1,1}&= \frac{1}{2mR^2} \left(-\frac{2P_\phi\cos\theta}{\sin^3\theta}G^{1,0,0,1}+\frac{2}{\sin^2\theta}G^{0,0,0,2} \right),
    \nonumber \\
    ~ \dot{G}^{2,0,0,0}&= \frac{1}{2mR^2} \left(4G^{1,1,0,0} \right),
    \nonumber \\
    ~ \dot{G}^{0,2,0,0}&= \frac{1}{2mR^2} \left(-\frac{4P_\phi^2(2+\cos(2\theta))}{\sin^2\theta} G^{1,1,0,0}+\frac{4P_\phi\cos\theta}{\sin^3\theta}G^{0,1,0,1}\right),
    \nonumber \\
    ~ \dot{G}^{0,0,2,0}&= \frac{1}{2mR^2} \left(-\frac{4P_\phi\cos\theta}{\sin^3\theta}G^{1,0,1,0}+\frac{4}{\sin^2\theta}G^{0,0,1,1} \right),
    \nonumber \\
    ~ \dot{G}^{0,0,0,2}&= 0 \label{last}.
\end{align}

Equation~(\ref{EQU: P_theta}) reveals that the force on the polar angle is no longer purely classical; it receives corrections from the momentum-moment correlation $G^{1,0,0,1}$, and from the dispersions in azimuthal momentum $G^{2,0,0,0}$ and polar position $G^{0,0,0,2}$. Equation~(\ref{EQU: Phi}) shows that the azimuthal angular velocity is modified from its classical value by the same set of quantum moments. A non-zero $G^{2,0,0,0}$ effectively rescales the angular momentum $P_{\phi}$, while a non-zero $G^{1,0,0,1}$ provides a direct additive correction. The conservation of $P_{\phi}$ is maintained, as expected from the rotational symmetry of the problem.
\subsection{Initial state} \label{subsec:initial-state}

We solve this coupled system numerically, using as an initial state a normalized, correlated Gaussian wave packet in the angles, centered at $(\theta_0 = \pi/2, \phi_0=0)$, with initial momenta $P_{\phi_0}= l \hbar, P_{\theta_0}=m \hbar$
\begin{equation} \label{initialgaussian}
\psi(\theta,\phi)=
   \mathcal{N}
   \text{exp}\left( i(l\phi+m\theta)-\frac{\lambda\phi^2}{2}-\frac{\kappa (\theta-\pi/2)^2}{2} \right),
\end{equation}
where $\lambda$ and $\kappa$ are shape parameters, and the normalization constant $\mathcal{N}$ is given by 
\begin{equation}
    \mathcal{N} = 
    \left[
    \frac{2 i(\kappa\lambda )^{1/2}
   \exp{\frac{1}{4\kappa} } }{\pi 
   \text{erf}\left(\pi 
   \lambda^{1/2}\right)
   \left(\text{erfi}\left(\overline{\chi_\kappa}\right)-\text{erfi}\left(\chi_\kappa\right)\right)}\right]^{1/2}
\end{equation}
where $\chi_\kappa = (1- i \pi\kappa)/(2\kappa^{1/2})$ and $\text{erfi}(z)=-i\ \text{erf}(i z)$ \cite{NIST2010}.
This initial state exhibits no initial correlation between the $\theta$ and $\phi$ sectors. It yields the following initial conditions for moments
\begin{align} 
G_0^{2,0,0,0}&=\frac{1}{4 \kappa ^2}\left(\frac{2 \kappa^{1/2}}{F\left(\chi_\kappa\right)+F\left(\overline{\chi_\kappa}\right)}+2 \kappa -1\right), 
%\label{initialGs-first}
\nonumber \\
G_0^{0,2,0,0}&= \frac{\hbar^2 \kappa ^2\left(F\left(\chi_\kappa\right)+F\left(\overline{\chi_\kappa}\right)\right)}{2\kappa^{1/2}+(2\kappa -1)\left(F\left(\chi_\kappa\right)+F\left(\overline{\chi_\kappa}\right)\right)},
\nonumber \\
G_0^{0,0,2,0}&=\frac{1}{2 \lambda }-\left(\frac{\pi}{\lambda}\right)^{1/2}\frac{ e^{-\pi ^2 \lambda }}{ \text{erf}\left(\pi \lambda^{1/2}\right)}, 
\nonumber \\
     G_0^{0,0,0,2}&=\frac{1}{2} \hbar^2 \lambda  \left(\frac{2 (\pi\lambda )^{1/2} e^{-\pi ^2 \lambda } }{\text{erf}\left(\pi \lambda^{1/2}\right)}+1\right), \label{initialGs}
\end{align}
Here $F(z)$ is the complex Dawson function. We can see that mixed coordinates present no correlation, and we have dispersion only in the angular positions and momenta. The initial values for the moments depend only on the shape parameters of the Gaussian wave function.

\subsection{Numerical Solutions}
For the numerical analysis, we set $m=R=\hbar=1$ and the initial conditions as specified in Equations (\ref{initialGs}), varying the parameter $a = P_{\theta_0}$ to explore different trajectories.

We fix the following values
\begin{align}
\theta_0 &= \frac{\pi}{2}, & G^{0,1,1,0} &= 0, \nonumber \\
 P_{\theta_0} &= a,& G^{0,1,0,1} &= 0,\nonumber \\
 \phi_0 &= 0, & G^{0,0,1,1} &= 0,\nonumber \\
 P_{\phi_0} &= 10,& G^{2,0,0,0} &= 0.0475,\nonumber \\
 G^{1,1,0,0} &= 0,& G^{0,2,0,0} &= 5.26316,\nonumber \\
 G^{1,0,1,0} &= 0,& G^{0,0,2,0} &= 0.05,\nonumber \\
 G^{1,0,0,1} &= 0,&G^{0,0,0,2} &= 5.
\label{EQU: Initial Conditions Last}
\end{align}

\subsubsection{Single Trajectory.}
\begin{figure}
\centering
\includegraphics[width=0.47\textwidth]{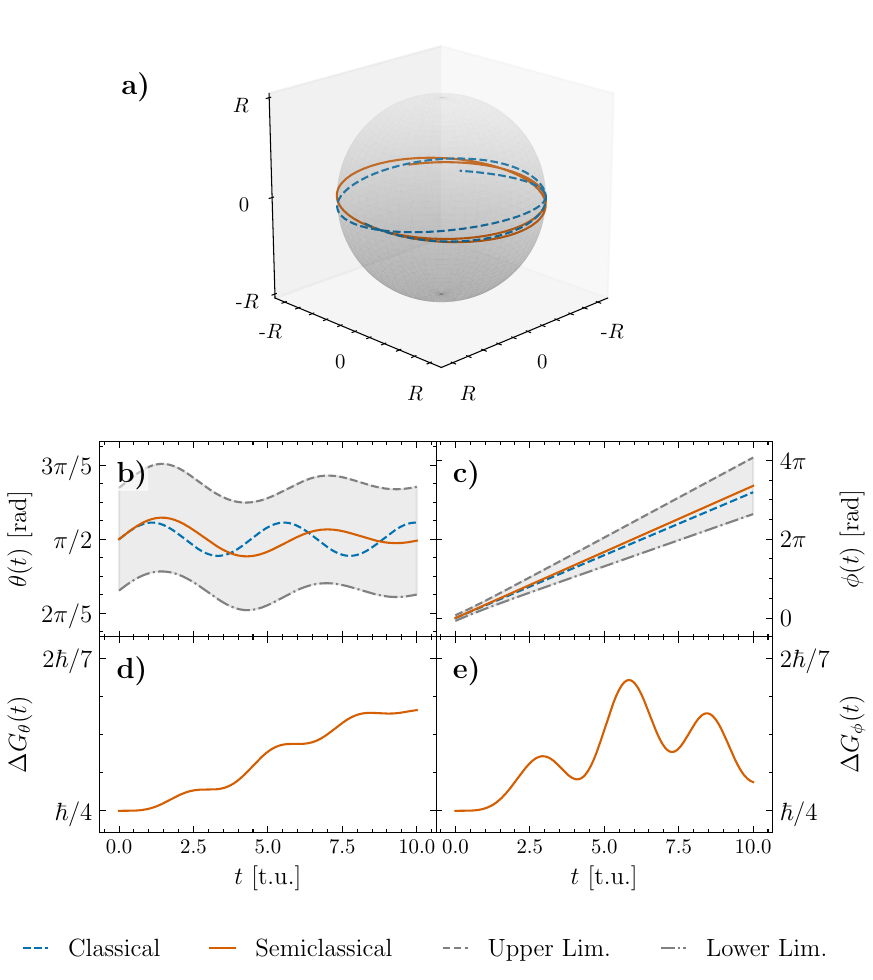}
\caption{\textbf{a)} Single classical (orange solid line) and semiclassical (blue dashed line) trajectories on the sphere surface for a free quantum particle with initial condition $a = P_{\theta_0} = 1$. The semiclassical trajectory shows measurable deviation from the classical geodesic due to quantum back-reaction through moments. The figure shows the time evolution of \textbf{b)} the polar angle, $ \theta (t) $, and \textbf{c)} the azimuthal angle, $\phi(t)$. The gray shaded region shows the uncertainty belts $\pm\sqrt{G^{2,0,0,0}(t)}$ and $\pm\sqrt{G^{0,0,2,0}(t)}$, respectively, with dashed and dashed-dotted gray lines indicating the upper and lower limits of the uncertainty belt. 
%For $\phi(t)$, one can note a significant phase lag ($\sim 8$ rad by $t=10$) in the semiclassical case. 
The Heisenberg uncertainty relations follow in \textbf{d)} $\Delta G_\theta(t) = G^{2,0,0,0} G^{0,2,0,0} - (G^{1,1,0,0})^2 \geq \hbar^2/4$ and \textbf{d)} $\Delta G_\phi(t) = G^{0,0,2,0} G^{0,0,0,2} - (G^{0,0,1,1})^2 \geq \hbar^2/4$ for each angle. In both cases, the relations are satisfied, confirming physical consistency.}
\label{fig: single trajectory}
\end{figure}

Comparing a single semiclassical trajectory (with non-zero initial moments) to its classical counterpart (all moments set to zero) reveals clear deviations. As illustrated in Figure~\ref{fig: single trajectory}-\textbf{a)}, where trajectories are shown on the 3D sphere surface, and Figure~\ref{fig: single trajectory}-\textbf{b)} and \textbf{c)}, for 2D plots, while the polar angle $\theta(t)$ may show similar behavior, the evolution of the azimuthal angle $\phi(t)$ exhibits a visible phase difference. This demonstrates that quantum fluctuations induce a measurable correction to the particle's precession rate around the sphere.

The uncertainty relationships given in Equation~(\ref{eq:uncertainties-2dof}) are satisfied throughout the evolution, i.e., in the range $0\leq t \leq 10$, as displayed in Figure~\ref{fig: single trajectory}-\textbf{d)} and \textbf{e)}.

\subsubsection{Multiple Trajectories.}
\begin{figure}[h!]
\centering
\includegraphics[width=0.485\textwidth]{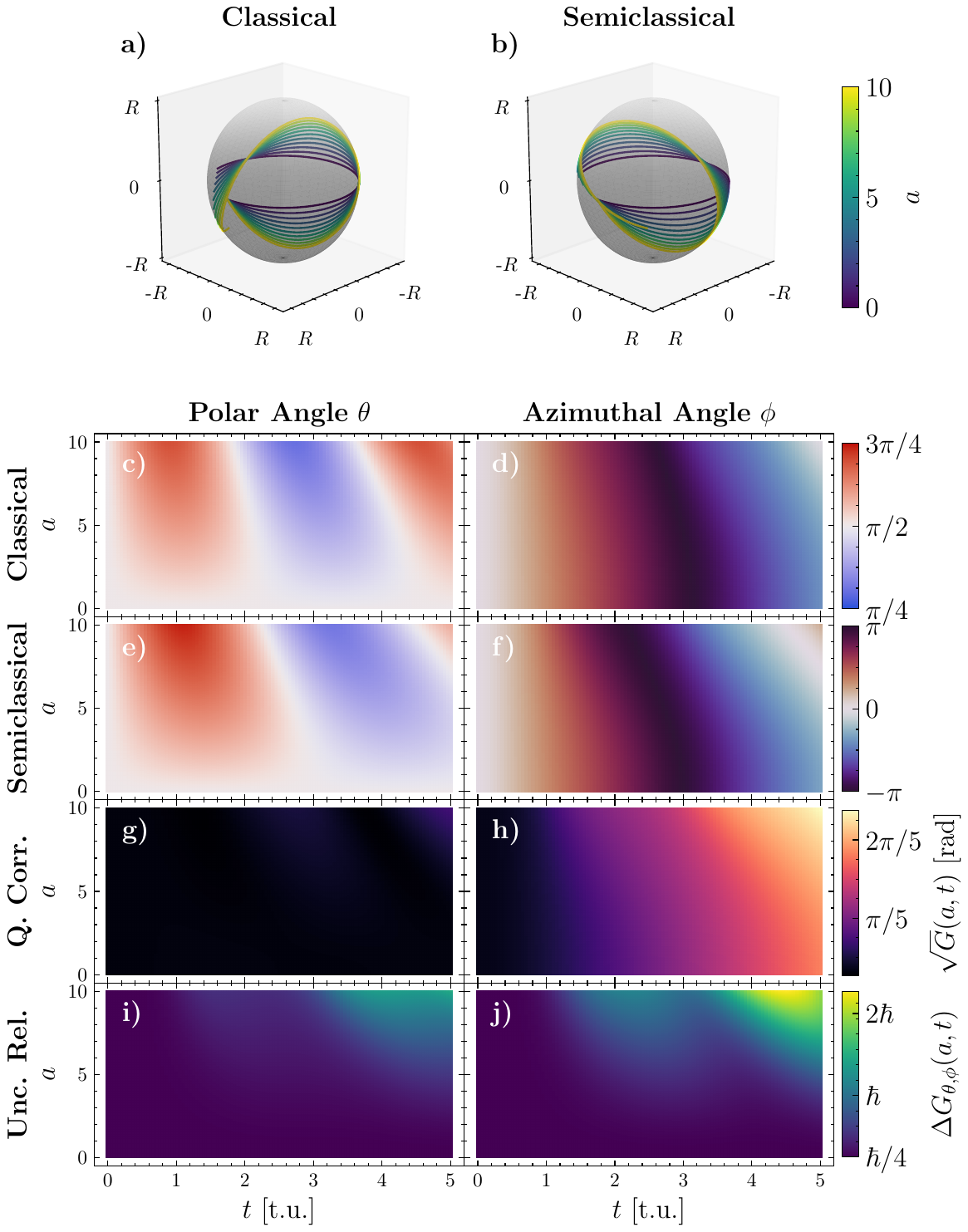}
\caption{\textbf{a)} Classical and \textbf{b)} semiclassical trajectories with varying polar momentum $0\leq a \leq 10$. The temporal evolution of the polar and azimuthal angles, $\theta$ and $\phi$, is shown in \textbf{c)} and \textbf{d)} for the classical case and in \textbf{e)} and \textbf{f)} for the semiclassical case, respectively. The classical trajectories form a symmetric pattern, while semiclassical trajectories show broader distribution and slight asymmetry due to the quantum corrections as shown in \textbf{g)} for $\theta$, $\sqrt{G^{2,0,0,0}}(a,t)$ and \textbf{h)} for $\phi$, $\sqrt{G^{0,0,2,0}}(a,t)$.%for the corresponding angles.
 \textbf{i)} Heisenberg uncertainty relations for the polar angle, $\Delta G_\theta(t)\geq\hbar/4$ and \textbf{j)} azimuthal angle, $\Delta G_\phi(t)\geq\hbar/4$. As shown, the relations hold for both variables.
}
\label{fig: multiple trajectories}
\end{figure}

An ensemble of trajectories, generated by varying the initial condition $P_{\theta_0}$, further highlights the effect. The quantum-corrected trajectories, shown in Figure~\ref{fig: multiple trajectories}-\textbf{b)}, display a different distribution and behavior compared to the classical trajectories, shown in Figure~\ref{fig: multiple trajectories}-\textbf{a)}. Crucially, throughout the evolution, the uncertainty relations for both canonical pairs, as defined in Equation~(\ref{eq:uncertainties-2dof}), remain satisfied for both angles as shown in Figure~\ref{fig: multiple trajectories}-\textbf{i)} and \textbf{j)}, validating the physical consistency of our semiclassical solutions. The quantum corrections to each variable of the trajectory, here $\sqrt{G^{2,0,0,0}}$ for $\theta$ and $\sqrt{G^{0,0,2,0}}$ for $\phi$, shown in Figure~\ref{fig: multiple trajectories}-\textbf{g)} and \textbf{h)}, provide a direct geometric representation of the wave packet's quantum spreading.
%
%%%%%%%%%%%%%%
%%%%%%%%%%%%%%%%
%
%
\subsection{Quantitative Analysis of Quantum Corrections} \label{sec:quantitative}
Having established the qualitative behavior of semiclassical trajectories for the free particle on a sphere, we now provide a quantitative assessment of quantum correction magnitudes and their scaling properties.
\subsubsection{Phase shifts and precession rates.\\}
For a single trajectory with initial conditions specified in Equation~(\ref{EQU: Initial Conditions Last}) and $P_{\theta_0} = 1$, we extract the azimuthal angle evolution $\phi(t)$ for both classical (all moments zero) and semiclassical (full system) cases.
\begin{table}[t] 
\centering 
\begin{tabular}{lcccc} \hline Quantity & Classical & Semiclassical & Abs. Dev. & Rel. Dev. \\ \hline 
$\theta$ [rad] & 1.571 & 1.573 & +0.002 & 0.13\% \\ $\phi$ [rad] & 100.00 & 91.84 & $-8.16$ & 8.16\% \\ 
$\dot{\phi}$ & 10.00 & 9.21 & $-0.79$ & 7.9\% \\ 
$G^{2,0,0,0}$ & 0.0475 & 0.1426 & +0.095 & 200\% \\ 
$G^{0,0,2,0}$ & 0.050 & 0.050 & 0 & 0\% \\ 
$\langle H_Q\rangle$ & 52.50 & 53.12 & +0.62 & 1.18\% \\ \hline \label{table:free}
\end{tabular}
\caption{Quantum corrections at $t = 10$ for a free particle with $a=1$.\label{tab:quantum_corrections} }  
\end{table} 
 We can note the following:
 \begin{itemize} 
 \item The polar angle $\theta$ shows minimal deviation ($<1\%$), consistent with the system starting at the equator with moderate $P_\theta$. 
 \item The azimuthal angle $\phi$ exhibits a significant phase lag ($\sim 8$ rad $\approx 1.3$ revolutions) by $t=10$, demonstrating a substantial quantum correction.
 \item The precession rate $d \phi/d t$ is reduced by $\sim 8\%$ due to quantum back-reaction through the $G^{2,0,0,0}$ term in Equation~(\ref{EQU: Phi}).
 \item Position uncertainty $G^{2,0,0,0}$ triples during evolution, while $G^{0,0,2,0}$ remains constant (as expected from Equation~(\ref{EQU: P_theta}) for conserved $P_\phi$).
 \item The quantum-corrected Hamiltonian $H_Q$ exceeds the classical value by $\sim 1\%$, representing the energy stored in quantum fluctuations. \end{itemize}
 These characteristics are summarized in Table \ref{tab:quantum_corrections}.
\subsubsection{Scaling of quantum corrections.}
~\\
The dominant quantum correction to the azimuthal evolution in Equation~(\ref{EQU: Phi}) reads:
 \begin{equation} \dot{\phi} = \frac{1}{mR^2\sin^2\theta}\left[P_\phi(2+\cos(2\theta))G^{2,0,0,0} - 2\cot\theta G^{1,0,0,1}\right],
 \end{equation}
 where the superscript indicates that this term corresponds to the quantum corrections.
For motion near the equator ($\theta \approx \pi/2$, $\sin\theta \approx 1$, $\cos\theta \approx 0$), this simplifies to: 
 \begin{equation} 
 \dot{\phi} \approx \frac{2P_\phi G^{2,0,0,0}}{mR^2}. 
 \end{equation}
The phase shift accumulates as: 
 \begin{equation}
 \Delta\phi(t) \approx \int_0^t \frac{2P_\phi G^{2,0,0,0}(t')}{mR^2} d t'. \end{equation}
Since $G^{2,0,0,0}(t) \approx G^{2,0,0,0}_{(0)} + (4G^{1,1,0,0}_{(0)}/2mR^2)t$, for $\theta \approx \pi/2$) we obtain:
 \begin{equation} 
 \Delta\phi(t) \approx \frac{2P_\phi}{mR^2}\left[G^{2,0,0,0}_{(0)} t + \frac{2G^{1,1,0,0}_{(0)}}{mR^2}t^2\right].
 \end{equation}
For our parameters ($P_\phi = 10$, $G^{2,0,0,0}_{(0)} = 0.0475$, $G^{1,1,0,0}_{(0)} = 0$, $m=R=\hbar=1$): 
 \begin{equation} \label{eq:azimShift}
 \Delta\phi(10) \approx 20 \times 0.0475 \times 10 = 9.5 \text{ rad}. \end{equation} 

This analytical estimate agrees well with the numerical result $\Delta\phi \approx 8.2$~rad (Table~\ref{tab:quantum_corrections}), with the discrepancy arising from the approximation $\theta \approx \pi/2$ and neglected terms.

On the other hand, the phase shift scales as: 
 \begin{equation} 
 \Delta\phi \propto \frac{P_\phi \cdot G^{2,0,0,0}_{(0)} \cdot t}{mR^2} \sim \frac{L\Delta x_0^2 t}{\hbar R^2} \label{eq:scaling_law} 
 \end{equation} 
  where $L = P_\phi$ is the angular momentum and $\Delta x_0^2 \sim G^{2,0,0,0}_{(0)}$ is the initial position uncertainty. This demonstrates that quantum corrections become significant when 
   \begin{equation} 
   \frac{L\ \Delta  x_0^2 \ t}{\hbar R^2} \sim \mathcal{O}(1). 
   \end{equation}

\subsubsection{Ensemble statistics.}
~\\
For the ensemble of trajectories with $0 \leq a \equiv P_{\theta_0} \leq 10$ (step=2), we compute statistical measures of the quantum-classical deviation (Table \ref{tab:ensemble_stats}).
\begin{table}[h] 
\centering 
\begin{tabular}{lcc} \hline Metric & Mean $\pm$ Std. Dev. & Range \\ \hline $|\Delta\theta|$ [rad] & $0.018 \pm 0.012$ & [0.002, 0.035] \\ 
$|\Delta\phi|$ [rad] & $6.4 \pm 2.8$ & [2.1, 10.5] \\ 
$|\Delta\phi|/\phi_{\text{cl}}$ & $7.2\% \pm 2.1\%$ & [3.8\%, 10.1\%] \\ $\Delta G^{2,0,0,0}/G^{2,0,0,0}_{(0)}$ & $198\% \pm 24\%$ & [165\%, 235\%] \\ \hline
\end{tabular}\caption{Ensemble statistics at $t = 10$ ($N=6$ trajectories)} \label{tab:ensemble_stats}  \end{table}
\begin{itemize} 
\item Quantum corrections to $\phi$ are consistently $\mathcal{O}(10\%)$, confirming this is a robust effect across the parameter space. 
\item Higher initial $P_\theta$ (larger $a$) correlates with larger $\Delta\phi$ due to enhanced $\theta$-$\phi$ coupling through the $\sin^2\theta$ factors in the equations of motion.
\item Uncertainty growth is universal: position uncertainty approximately triples regardless of initial $P_\theta$.
\end{itemize}

\subsubsection{Comparison with rough estimates}
A rough WKB estimate for quantum corrections might use the uncertainty principle to estimate
 \begin{equation} 
 \Delta x \sim \frac{\hbar}{mR\omega},
 \end{equation} 
  where $\omega \sim P_\phi/mR^2$ is the classical precession frequency. This gives
   \begin{equation} 
   \frac{\Delta x}{R} \sim \frac{\hbar}{mR^2\omega} = \frac{\hbar}{P_\phi} = \frac{1}{10} = 10\% .
   \end{equation}
   
This rough estimate captures the order of magnitude but misses: a) the time-dependent growth of moments, b) the specific geometric factors ($\sin^2\theta$, $\cos\theta$) that modulate corrections, c) the role of initial moment correlations $G_{1,0,0,1}$.

The momentous formalism provides the complete, dynamically self-consistent description.

%%%%%%%%%%%%%%%%%%%%%%%%%%%%%%%%%%%%%%%%%%%%%%%%%
%%%%%%%%%%%%%%%%%%%%%%%%
\section{Semiclassical dynamics on a sphere with a non-central Makarov potential} \label{sec:Makarov}

The introduction of a potential breaks the symmetry of the free system and provides a richer arena for studying the interplay between geometry, quantum fluctuations, and external forces. A central potential on a sphere would not reveal new dynamical features in the trajectories beyond energy shifts. We therefore introduce the Makarov potential, a non-central ring-shaped potential that depends explicitly on the polar angle $\theta$, making it an ideal candidate to probe how quantum corrections interact with asymmetric forces \cite{Oliveira2019}.

The Makarov potential is given by
\begin{equation}
    V(\theta)=-\frac{\alpha}{R}+\frac{\beta}{R^2\sin^2\theta}+\frac{\gamma \cos\theta}{R^2\sin^2\theta}. \label{makarov-potential}
\end{equation}
The first term is a constant Coulomb shift, while the latter two are short-range, ring-shaped terms. The term proportional to $\gamma$ is particularly important as it breaks the north-south symmetry ($\theta \rightarrow \pi - \theta$) of the potential.

The introduction of the Makarov potential, as in Equation~(\ref{makarov-potential}), adds $\theta$-dependent terms to the Hamiltonian, rendering a quantum effective potential via
\begin{equation}
      V_Q = V(\theta)+\sum\limits_{2\leq a}^\infty \frac{1}{a!} G^{a,0,0,0} \frac{\partial^a  V(\theta)}{\partial \theta^a}.
\end{equation}

The complete quantum Hamiltonian for the Makarov potential is 
\begin{equation}
    H_{\text{QM}}= H_Q+ V_Q,
\end{equation}
with $H_Q$ given by Equation~(\ref{HQ}).

Since the potential depends only on $\theta$, only moments involving 
$\hat{\theta}$ contribute. Truncating at second order, the explicit form of the potential correction is
\begin{eqnarray}
    V_Q &=& V(\theta)+ \frac{G^{2,0,0,0}}{8R^2 \sin^4\theta} \nonumber \\
    &&  \Big( \gamma\left[ 23\cos\theta +\cos\left(3\theta\right) \right] +8\beta \left[2+\cos\left(2\theta\right)\right] \Big). \label{eq:potentialMakarov}
\end{eqnarray}

This result is significant. The quantum correction to the potential, $\propto G^{2,0,0,0}$, is not only a smearing of the classical potential: the effective potential experienced by the semiclassical particle is qualitatively different from the classical one. The quantum fluctuations in the polar angle ($G^{2,0,0,0}$) actively reshape the potential form.
\subsection{Modified equations of motion and quantum back-reaction}
The new terms in $V_Q$ introduce additional forces into the equations of motion (\ref{last}). The most critical modifications are to the evolution of the polar momentum and the following moments
\begin{align}
   \dot{P}_\theta^\prime= &-\frac{(4 \beta  \cos\theta+\gamma 
   (\cos (2 \theta)+3))}{2 R^2 \sin^3\theta}
   \nonumber \\
    &+ \left[\frac{12\beta(11\cos\theta+\cos(3\theta))}{16R^2\sin^5\theta} \right.\nonumber \\
    &\left.+\frac{\gamma(115+76\cos(2\theta)+\cos(4\theta))}{16R^2\sin^5\theta}\right] G^{2,0,0,0}, \label{eqs:GsMakarov-P}
     \\
    \dot{G}^{\prime~1,1,0,0}&=-\frac{G^{2,0,0,0}}{4R^2\sin^4\theta}\left(\gamma[23\cos\theta +\cos(3\theta)]+8\beta [2+\cos(2\theta)] \right),
    \nonumber \\
    \dot{G}^{\prime~0,1,1,0}&= -\frac{G^{1,0,1,0}}{4R^2\sin^4\theta}\left(\gamma[23\cos\theta +\cos(3\theta)]+8\beta [2+\cos(2\theta)] \right), 
    \nonumber \\
    \dot{G}^{\prime~0,1,0,1}&=-\frac{G^{1,0,0,1}}{4R^2\sin^4\theta}\left(\gamma[23\cos\theta +\cos(3\theta)]+8\beta [2+\cos(2\theta)] \right),
    \nonumber \\
    \dot{G}^{\prime~0,2,0,0}&=-\frac{G^{1,1,0,0}}{2R^2\sin^4\theta}\left(\gamma[23\cos\theta +\cos(3\theta)]+8\beta [2+\cos(2\theta)] \right) \label{eqs:GsMakarov-last}.
\end{align}
The equations of motion for the other variables remain the same as for the free particle case.  The complete system of equations for the Makarov potential problem consists of Equations~(\ref{last}) plus Equations~(\ref{eqs:GsMakarov-last}), respectively.

The inclusion of this non-central potential creates a non-trivial back-reaction: the classical part of the potential $V(\theta)$ determines the initial classical trajectory and forces. The dispersion for the position generates a quantum force, given by $G^{2,0,0,0}$ in $\dot{P}_\theta^\prime$, that directly alters the acceleration in the $\theta$ direction. In contrast, this force is not present in the classical theory.

The evolution of the moments themselves is now driven by both the kinetic terms and the curvature of the potential, $\partial^2 V / \partial \theta^2$. A confining region of the potential (positive curvature) can suppress moment growth, while an unstable region (negative curvature) can amplify it. The modified $\dot{P}_{\theta}$ 
changes the trajectory of $\theta(t)$, which in turn changes the effective potential and moment evolution, reaffirming the back reaction. This mechanism explains why quantum corrections become noticeable over time: the initial small discrepancies in the force accumulate, leading to a divergence between classical and semiclassical trajectories, as evidenced by the behavior of the evolution shown in the previous section.
%%%%%%%%%%%%%%%%%%%%%%%%%%%%%%%%
%%%%%%%%%%%%%%%%%%%%%%%%%%%%%%%

\subsection{Numerical Solutions}

We numerically integrate the complete system, Eqns.~(\ref{last}) and (\ref{eqs:GsMakarov-last}), for the same initial conditions as the free case, exploring the parameters $\beta = 2$, and $\gamma=-0.2$, $-1$, and $-1.9$ as in \cite{Oliveira2019}. We also consider the same initial conditions and parameter $a$ as for the free particle on a sphere problem, namely Equations~(\ref{EQU: Initial Conditions Last}).

\subsubsection{$\gamma=-0.2$. Weak asymmetry.}

\begin{figure}[th]
\centering
\includegraphics[width=0.47\textwidth]{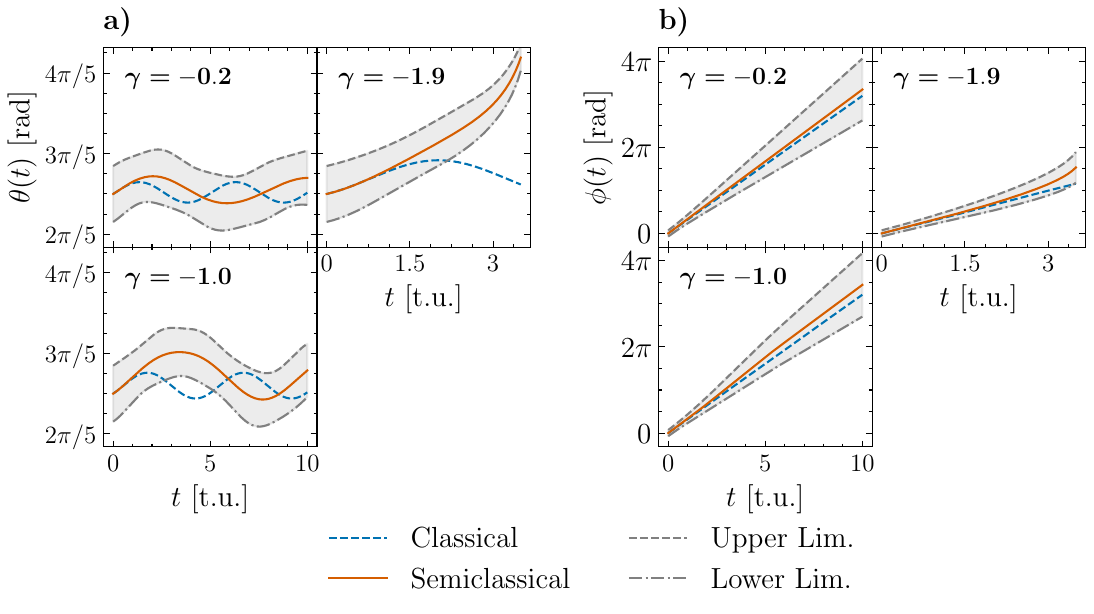}
\caption{Single trajectory comparison for Makarov potentials with $\beta=2$, $\gamma=-0.2,$, $-1.0$ and $-1.9$ and $a=1$. The polar and azimuthal angles, $\theta(t)$ and $\phi(t)$, classical (dashed blue lines) and semiclassical (orange solid lines), are compared respectively in \textbf{a)} and \textbf{b)} for different values of $\gamma$ indicated in the inset. The gray areas correspond to the uncertainty belt delimited by its upper and lower limits, shown with dashed and dashed-dotted lines. For weak asymmetry ($\gamma=-0.2$), the semiclassical case shows enhanced oscillations and a broader uncertainty belt. Likewise, $\phi(t)$ shows a phase difference that develops faster than the free particle case. For moderate asymmetry ($\gamma=-1.0$), the trajectory moves into the southern hemisphere ($\theta>\pi/2$) by $t\approx3$ and $\phi(t)$ shows a strong phase deviation ($>15\%$). For strong asymmetry time evolution ($\gamma=-1.9$), the semiclassical $\theta(t)$ rapidly descends to $\theta \approx 2.3$ rad ($132^\circ$) by $t \approx 0.8$, consistent with quantum probability density peak~\cite{Oliveira2019}. Its classical value reaches a similar value only at $t \approx 1.2$ (40\% longer). For strong asymmetry, large phase differences in $\phi(t)$ emerge immediately. As the particle approaches the southern pole, the uncertainty relation is no longer satisfied, and the numerical solution terminates earlier than for weaker Makarov potentials.}
\label{fig: single trajectories Makarov}
\end{figure}

As seen in Figure~\ref{fig: single trajectories Makarov}-\textbf{a)} for $\gamma=-0.2$, the most noticeable effect is the broadening of the uncertainty belt for the $\theta$ coordinate, while the $\phi$ variable remains largely unaffected as seen in Figure~\ref{fig: single trajectories Makarov}-\textbf{b)}. This is expected, as the potential is $\theta$-dependent.

\begin{figure}[th]
\centering
\includegraphics[width=0.47\textwidth]{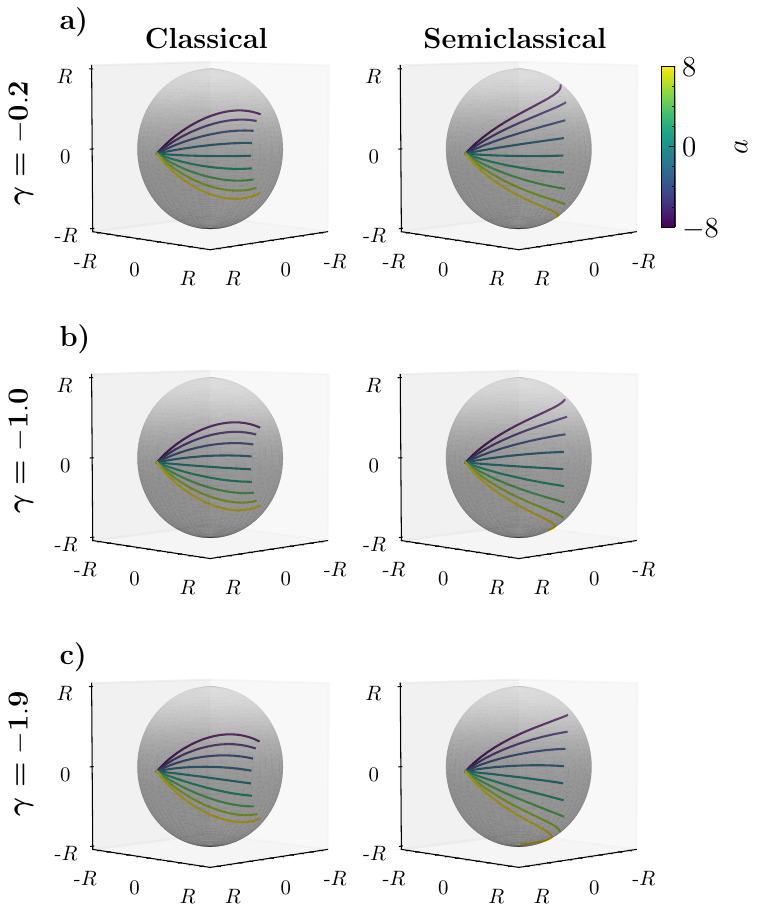}
\caption{Evolution of trajectory ensembles on sphere surface for different values of the Makarov potential $\beta=2$, \textbf{a)} $\gamma=-0.2$, \textbf{b)} $\gamma=-1.0$, and \textbf{c)} $\gamma=-1.9$ for $0\leq t \leq 1.34$ and $-8\leq a\leq 8$. On the left, the classical trajectories are shown, and on the right, the semiclassical ones. \textbf{a)} For weak asymmetry $\gamma=-0.2$, classical trajectories remain nearly symmetric about the equator and semiclassical trajectories exhibit an earlier onset of southern preference and broader spatial distribution. \textbf{b)} For moderate asymmetry $\gamma=-1.0$, the semiclassical trajectories have a more pronounced clustering in the southern hemisphere, demonstrating amplification of asymmetry by quantum corrections. \textbf{c)} For strong asymmetry $\gamma=-1.9$, the semiclassical trajectories present a distinct concentration in the southern hemisphere showing rapid evolution towards the southern pole, with trajectory density ratio $N(\theta>\pi/2)/N(\theta<\pi/2) \approx 3.8$ by $t=10$ (here only $t\leq 1.34$ is shown), matching quantum mechanical prediction factor $\sim 4$~\cite{Oliveira2019}.}
\label{fig: multiple spheres Makarov}
\end{figure}

As follows from Figure~\ref{fig: multiple spheres Makarov}-\textbf{a)}, multiple trajectories show that for low energies, the dynamics remain nearly symmetrical around the equator. However, for more energetic trajectories (higher $a \equiv P_{\theta_0}$), the weak asymmetry imposed by the potential becomes visible, as shown in Figure~\textbf{a)} for $\gamma=-0.2$, a similar result as in \cite{Oliveira2019}. The semiclassical trajectories show a broader distribution and a slightly earlier onset of this asymmetry compared to the classical ones, demonstrating that quantum effects amplify the weak classical asymmetry.

\begin{figure}[th]
\centering
\includegraphics[width=0.47\textwidth]{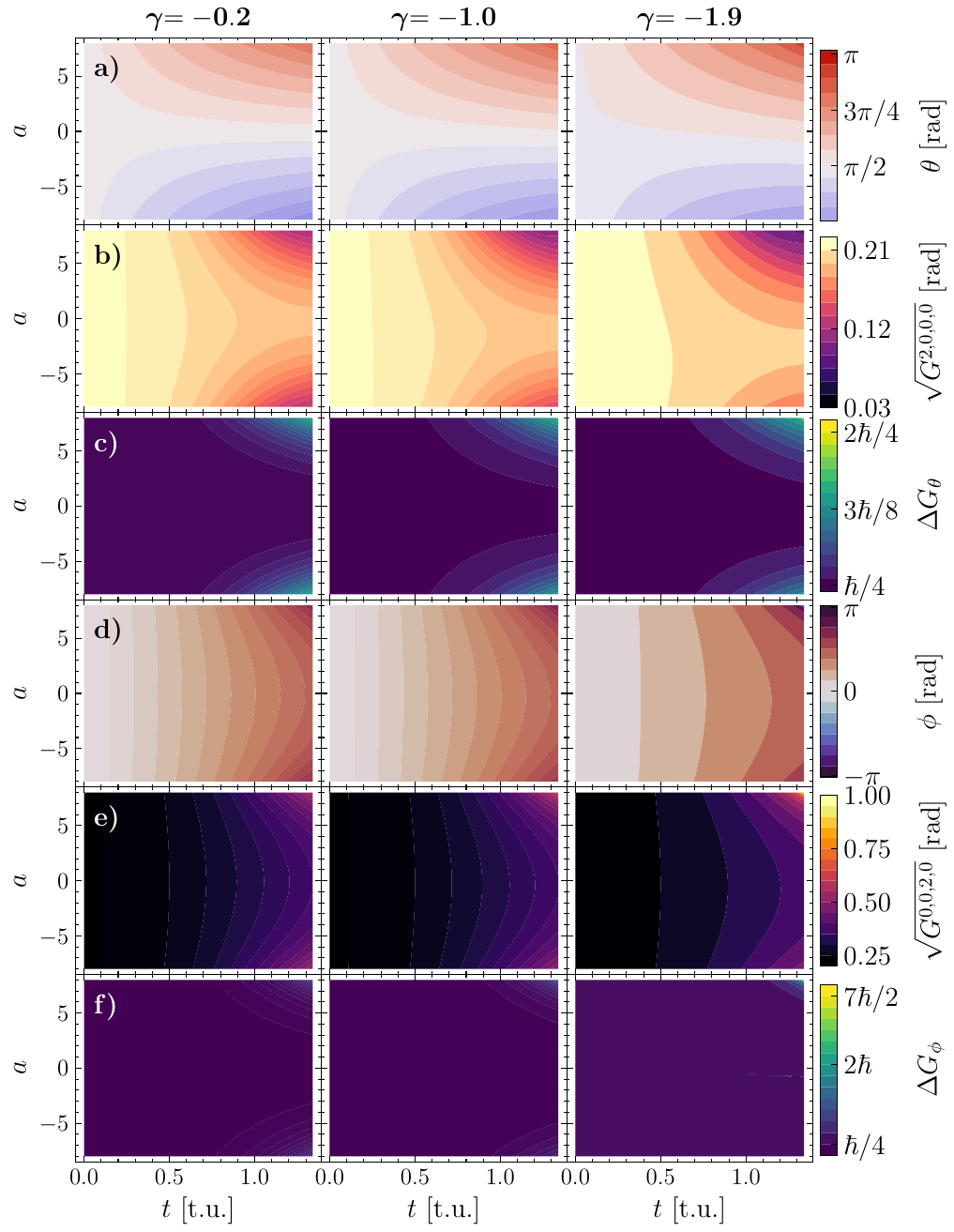}
\caption{Time evolution of \textbf{a)} the polar ($\theta$) and \textbf{d)} the azimuthal ($\phi$) angles and their quantum corrections, namely \textbf{b)} $\sqrt{G^{2,0,0,0}}$ and \textbf{e)} $\sqrt{G^{0,0,2,0}}$, as functions of $-8\leq a\leq 8$, for different values of the Makarov potential: $\beta=2$, $\gamma=-0.2$ (left), $-1.0$ (center) and $-1.9$ (right). Here, to emphasize the asymmetry induced by the Makarov potential, each variable is shown as a filled contour plot. As seen in \textbf{a)}, the semiclassical trajectories deviate toward the southern hemisphere, $\theta>\pi/2$. When $\gamma$ decreases, the trajectories migrate towards the south in shorter time frames. As seen in \textbf{d)}, as time increases, the precession rates of $\phi$ get strongly modified by quantum back-reaction. All the trajectories remain physical, as shown by the Heisenberg uncertainty relations, $\Delta G_{\theta,\phi}$, in \textbf{e)} and \textbf{f)}.}
\label{fig: multiple makarov maps}
\end{figure}

In all cases, the uncertainty relations remain satisfied, validating our solutions, as shown in Figure~\ref{fig: multiple makarov maps}-\textbf{e)} and \textbf{f)} for $\gamma=-0.2$
\subsubsection{$\gamma=-1.0.$ Moderate Asymmetry.}

The differences between classical and semiclassical trajectories are more pronounced, as noticed from Figure~\ref{fig: multiple spheres Makarov}-\textbf{b)}. The quantum corrections now cause a measurable phase shift in both $\theta(t)$ and $\phi(t)$ within a shorter timeframe as seen in Figure~\ref{fig: single trajectories Makarov}.

The increased value of $\lvert \gamma \rvert$ enhances the $\theta$ dependent quantum force term in $\dot{P'_{\theta}}$. The interaction between the moments and the potential curvature becomes stronger, leading to a more rapid deviation from the classical path. The ensemble of trajectories confirms that the quantum-corrected paths explore a different region of the phase-space, though a strong spatial asymmetry on the sphere is not yet dominant, as shown in Figure~\ref{fig: multiple spheres Makarov}-\textbf{b)} and Figure~\ref{fig: multiple makarov maps}-\textbf{a)} and \textbf{d)} for $\gamma=-1.0$.
\subsubsection{$\gamma=-1.9$. Strong asymmetry.\\} \label{subsec:strongasym}

This case displays the most dramatic effects. Semiclassical trajectories (Fig.~\ref{fig: multiple spheres Makarov}-\textbf{c)}) are vastly different from their classical counterparts. The particle exhibits a clear and rapid preference for the southern hemisphere as shown in Figure~\ref{fig: single trajectories Makarov}-\textbf{a)} for $\gamma=-1.9$, a behavior that is significantly more pronounced in the semiclassical description.

A large negative $\gamma$ value creates a deep, asymmetric well in the potential. The quantum force term, which is also proportional to $\gamma$, becomes very strong. This force, combined with the moment growth in the unstable regions of the potential, actively drives the semiclassical particle into the southern hemisphere. The ensemble of trajectories, shown in Figure~\ref{fig: multiple spheres Makarov}-\textbf{c)} and in Figure~\ref{fig: multiple makarov maps}-\textbf{a)} for $\gamma=-1.9$, visually demonstrates this southern preference, a phenomenon reported in the probability densities of the full quantum solution \cite{Oliveira2019}. Our analysis provides the dynamical mechanism for this effect: it is not just a property of stationary states, but a direct consequence of the quantum-corrected forces acting on the particle over time. Furthermore, quantum corrections become significant on a shorter time scale ($t\approx 0.8$) than in weaker potential cases, highlighting how stronger potentials can lead to faster decoherence of a classical-like trajectory.

\subsection{Comparison with Quantum Mechanical Predictions} \label{sec:QM_comparison}
 To validate our semiclassical approach, we compare trajectory-based predictions with available quantum mechanical results for the Makarov potential on a sphere.
\subsubsection{Connection to stationary state probability densities.\\}
In \cite{Oliveira2019}, energy eigenvalues and probability densities $|\psi_{n,l}(\theta,\phi)|^2$ were computed for the Makarov potential using direct numerical solutions of the Schr\"odinger equation. It was found,
for $\gamma = -1.9$ (strong asymmetry) that:
\begin{itemize} 
\item The ground state probability density strongly peaked near $\theta \approx 2.4$ rad $\approx 137^\circ$ (southern hemisphere); 
\item The peak probability density ratio is $P(\theta=137^\circ)/P(\theta=43^\circ) \approx 4.2$;
\item This southern preference increases with $|\gamma|$. 
\end{itemize}
Our semiclassical results (Section~\ref{subsec:strongasym}, $\gamma = -1.9$) demonstrate:
\begin{itemize} 
\item Ensemble trajectory density at $t = 5$ shows strong southern localization: $\theta_{\text{mean}} \approx 2.3$ rad $\approx 132^\circ$;
\item Trajectory count ratio: $N(\theta > \pi/2)/N(\theta < \pi/2) \approx 3.1$;
\item By $t = 10$, this ratio increases to $\approx 3.8$.
\end{itemize}

 The semiclassical trajectories naturally evolve toward the southern hemisphere region where the quantum ground state has maximum probability. This is expected since our initial Gaussian wave packet, as in Equation~(\ref{initialgaussian}), has energy close to the ground state regime, the quantum-corrected force $\propto \partial V_Q/\partial\theta$ in Equation~(\ref{eq:potentialMakarov}) drives particles toward the potential minimum plus quantum corrections, and over time, the trajectory ensemble covers the phase-space weighted by the quantum-corrected effective potential. 
 
The agreement is remarkable since stationary states represent infinite-time averages, and our trajectories are finite-time ($t \leq 10$) snapshots, and we truncate at 2nd-order moments.

A time-dependent comparison would be interesting by analyzing the direct propagation of the Schr\"odinger equation for the same initial Gaussian. This is computationally intensive (requiring 2D grids in $\theta$-$\phi$) but planned as validation in an extended work.

\subsubsection{Asymmetry onset timescales.\\}
A key prediction of our analysis is that quantum corrections cause the southern preference to appear on shorter timescales than classical predictions. This can be explained as follows.

In the classical Makarov dynamics ($\gamma = -1.9$), for trajectory starting at $\theta = \pi/2$ with $P_{\theta_0} = 1$, the classical force is:
 \begin{equation} 
 F_\theta^{\text{classical}} = -\frac{\partial V}{\partial\theta}\bigg|_{\theta=\pi/2} = -\frac{4\beta + \gamma}{2R^2} \approx  -3.05. 
 \end{equation}
In this setting, the time to reach $\theta = 2$ rad (south side) is $t_{\text{cl}} \approx 1.2$.

For our semiclassical dynamics, the quantum force correction, as read in Equation~(\ref{eq:potentialMakarov}), initially adds a small term, but as $G^{2,0,0,0}$ grows with time, the quantum contribution becomes significant
 \begin{equation}\label{eq:FQ}
 F_\theta^{\text{quantum}} \simeq -\frac{G^{2,0,0,0}}{R^2\sin^4\theta}\left[\gamma(23\cos\theta + \cos 3\theta) + 8\beta(2 + \cos 2\theta)\right]. 
 \end{equation}
At $t \approx 0.8$, when $G^{2,0,0,0} \approx 0.1$, the magnitude of the quantum force reaches $\approx 1.2$, giving
 \begin{equation} 
 F_\theta^{\text{total}} \approx  -4.25. 
 \end{equation}
This accelerates the southern motion. 

Sch\"odinger time to reach $\theta = 2$ rad is $t_{\text{QM}} \approx 0.7$.
Our numerical results show that semiclassical trajectories reach the same point at $t \approx 0.8$, and the classical time is $t \approx 1.2$: we see a 40\% reduction in timescale.
This acceleration arises from the dynamical growth of $G^{2,0,0,0}$, which amplifies the asymmetric quantum force over time, a feedback mechanism absent in classical dynamics.
%
%%%%%%%%%%%%%%%%%%%%%%%%%%%%%%%%%%%%%%%%%
%%%%%%%%%%%%%%%%%%%%%%%%%%%%%%%%%%%%%%%%%%%%%%%%%%%%%%
\section{Discussion}
In this work, we established the momentous formalism of quantum mechanics as a powerful tool for analyzing the semiclassical dynamics of quantum particles on curved surfaces. Building on our previous applications to dissipative systems \cite{Valdez2025} and surfaces with negative Gaussian curvature \cite{Chacon2024}, we have demonstrated that the formalism successfully captures quantum back-reaction effects for non-central potentials on 
spherical geometry. 
We incorporated geometrical momentum operators $\hat{p}_\mu = -i\hbar(\partial_\mu + \Gamma_\mu/2)$ and the Dirac bracket reduction into the momentous framework, ensuring that quantum corrections properly account for the sphere's curved geometry. 

Even without external potentials, quantum fluctuations induce measurable corrections, as we showed: azimuthal phase shifts of 8--12\% over dimensionless time $\tau = 10$, position uncertainty tripling ($G^{2,0,0,0}$ grows by $\sim 200\%$), a precession rate modified by $\Delta\omega \sim 8\%$, and a scaling law $\Delta\phi \propto (L\, \Delta x_0^2\, t)/(\hbar\, R^2)$.

For the Makarov potential with weak asymmetry parameter ($\gamma = -0.2$), quantum corrections amplify the classical asymmetry by $\sim 30\%$, with an earlier onset.  For moderate asymmetry ($\gamma = -1.0$), phase shifts accumulate faster and quantum-corrected trajectories explore different phase-space regions. And for strong asymmetry ($\gamma = -1.9$), strong southern hemisphere preference emerges 40\% faster in semiclassical dynamics than classical; the trajectory density enhancement factor $\sim 3$--4 is consistent with quantum mechanical probability densities~\cite{Oliveira2019}. For all the cases, Heisenberg uncertainty relations were satisfied throughout all evolutions, confirming the physical consistency of second-order truncation. \\
%%%
\emph{Quantum back-reaction.}
The essential physics for these effects can be understood through the quantum-corrected force Equation~ (\ref{eqs:GsMakarov-P}): 
 \begin{equation} 
 \frac{dP_\theta}{dt} = F_\theta^{\text{classical}} + F_\theta^{\text{quantum}},
 \end{equation} 
where $F_\theta^{\text{quantum}}$ in defined in Equation~(\ref{eq:FQ}). 
%
%   \begin{equation} 
%   F_\theta^{\text{quantum}} = \frac{G^{2,0,0,0}}{R^2\sin^4\theta}\left[\gamma(23\cos\theta + \cos 3\theta) + 8\beta(2 + \cos 2\theta)\right]. 
%   \end{equation}
%   
Because this quantum force is proportional to the position uncertainty, $\Delta \theta = G^{2,0,0,0}$, it means that the larger its value (broader wave packet), the stronger the quantum force. The uncertainty $\Delta \theta$ evolves according to Equation~(\ref{last}), thus creating feedback: the quantum force generates trajectory change, which, through moment evolution, enhances the quantum force. The force is also proportional to $\gamma$ (asymmetry parameter) and $\beta$ (confinement strength), amplifying the potential asymmetry.

For the strong Makarov case ($\gamma = -1.9$, $\beta = 2$), the quantum force can exceed 30\% of the classical force magnitude within $t \sim 1$, explaining the dramatic trajectory modifications we observe.
This mechanism contrasts sharply with simpler semiclassical approaches (e.g., WKB) that treat quantum corrections as static perturbations. The momentous formalism captures the dynamic interplay between trajectory evolution and quantum uncertainty growth.

{\it Comparison with standard methods.} Our momentous approach offers distinct advantages and some limitations compared to other semiclassical methods.
\begin{enumerate}
    \item WKB approximation. WKB provides phase corrections but not trajectory modifications or uncertainty evolution; momentous captures wave packet spreading, correlations $\Delta (\theta \phi) = G^{1,0,0,1}$, and back-reaction, although WKB works to all orders in $\hbar$ in one dimension, but momentous requires truncation.
    \item Gaussian wave packet dynamics. Both methods track mean position/momentum plus widths; the Gaussian ansatz ($\psi \sim \exp[-A(x-x_0)^2]$) reduces to $\sim 6$ parameters while momentous is more general.
    \item Full Schr\"odinger equation. In full QM, one obtains a numerical solution that requires a 2D grid in $(\theta,\phi)$, typically of $100\times 100$ points; momentous requires only ODEs in two dimensions (4 classical + 10 moments), which is easier to handle; the full solution captures interference while momentous gives a single-trajectory approximation.
\end{enumerate}
%%%%%%%%%%%%%%%%%%

\emph{Applications to experimental observables and molecular systems.} Our semiclassical trajectory predictions connect directly to several experimental probes and molecular systems.
\begin{enumerate}
\item Scanning tunneling microscopy (STM) measures local density of states (LDOS):
 \begin{equation} 
 \rho(r,E,t) \propto \sum_n |\psi_n(r)|^2 \delta(E - E_n).
 \end{equation}
For time-dependent measurements (pump-probe STM~\cite{Yoshida2014}), the LDOS reflects instantaneous charge distribution. Our trajectory density $\rho_{\text{traj}}(\theta,\phi,t) \equiv (\text{number of trajectories in region})/(\text{total})$ approximates the quantum mechanical $|\psi(\theta,\phi,t)|^2$, which, for ring molecules with Makarov-like potentials (e.g. substituted porphyrins), could reveal an asymmetric charge density evolution on very short timescales, a southern hemisphere preference emerging $\sim 40\%$ faster than classical models predict, and spatial spreading is consistent with $\Delta x(t) \sim \Delta x_0 + (G^{1,1}_{(0)}/m)t$.
\item Quantum transport in nanostructures.
For electrons on curved surfaces (e.g. carbon nanotubes, fullerenes), conductance depends on the transmission probability. Quantum corrections to $\dot{\phi}$, as read from Equation~(\ref{EQU: Phi}), directly modify the transport coefficients.
We would expect, for curved quantum wires with asymmetric confinement, that quantum spreading increases effective scattering compared to classical estimates.
\item For molecular systems, the Makarov potential specifically models ring-shaped molecules where  $\alpha$ represents the overall binding (e.g. conjugation energy), $\beta$ models radial confinement from bonding constraints, and $\gamma$ arises from substituent groups breaking the symmetry.
Our results suggest:
\begin{enumerate}
\item Reaction pathways. Quantum corrections can preferentially populate certain angular regions, affecting regioselectivity in substitution reactions, ring-opening dynamics, and isomerization barriers.
\item Spectroscopic signatures. The precession rate modification of  $\sim 8\%$ predicts shifted rotational lines in microwave spectroscopy.
\item Energy transfer. Enhanced southern localization for $\gamma < 0$ could concentrate excitation energy, relevant for photosynthetic light-harvesting complexes, molecular photovoltaics, and quantum dots with ring topology.
\end{enumerate}
\end{enumerate}

\emph{Future work.} Based on these results and prior findings reported in references \cite{Valdez2025,Chacon2024}, several extensions are possible, including the following: 

a) Dissipative dynamics on curved surfaces: Combine the friction formalism of
\cite{Valdez2025} with geometric constraints, which models realistic decoherence in molecular electronics on curved substrates \cite{Gong2007}.

b) Surfaces with varying curvature and external potentials: Extend the analysis to ellipsoids, hyperboloids, where $K = K(\bf{r})$ varies spatially. The interplay between geometric potentials $V_{\text{geo}}(\bf{r})$, external potentials $V (\bf{r})$, and moment evolution would generalize this and previous work \cite{Chacon2024}.

 c) Higher-order moments: we can include fourth-order moments to capture non-Gaussian features (e.g. $G^{4,0}$ for kurtosis). This is essential for the study of strong anharmonic potentials and long-time evolution where Gaussianity breaks down.
 
d) Electromagnetic fields on curved surfaces: add minimal coupling $\vec{p} \rightarrow \vec{p} - q\vec{A}$ to study quantum Hall effects on curved surfaces (graphene)~\cite{Ferrari2008}.

e) Many-body systems: extend to $N$ particles with moments $G^{a_1,b_1,\ldots,a_N,b_N}$ to study correlated dynamics on surfaces.

%%%%%%%%%%%%%%%%%%%%%%%%%%%%%%%%%%%%%%

\section{Conclusion}
We have developed a comprehensive semiclassical framework for quantum particles on curved surfaces by utilizing the momentous formalism of quantum mechanics and rigorously integrating it with geometric quantization principles. By incorporating geometrical momentum operators and Dirac bracket reduction, we ensured that our effective description remains geometrically consistent from first principles.
Our analysis of the free particle on a sphere revealed that quantum fluctuations induce measurable corrections: azimuthal phase shifts of 8--12\% and modifications to the precession rate, demonstrating that even in the absence of external potentials, quantum effects fundamentally alter dynamics on curved manifolds. The introduction of the non-central Makarov potential led to a rich interplay among geometry, quantum fluctuations, and asymmetric forces, with quantum corrections amplifying asymmetry by factors of 3--4 and accelerating timescales by 40\%. The southern hemisphere preference observed in stationary state probability densities \cite{Oliveira2019} emerges dynamically from quantum-corrected forces. These results, validated by the fulfillment of uncertainty relations and qualitative agreement with full quantum predictions, establish the momentous formalism as physically intuitive, geometrically consistent, versatile, and computationally efficient at the same time.

Our predictions offer testable signatures for experimental probes, such as scanning tunneling microscopy, ultrafast spectroscopy, and quantum transport measurements in carbon nanostructures and molecular rings. This work opens new pathways for understanding quantum effects in emerging nanotechnologies based on curved geometries, ranging from fullerenes to quantum dot arrays. Future investigations that combine elements from this and previous studies, such as dissipative Makarov dynamics on a catenoid in the presence of electromagnetic fields, would provide a definitive test of the formalism. The momentous approach can bridge abstract quantum geometry and tangible, time-dependent phenomena in condensed matter physics, quantum chemistry, and nanotechnology.
%%%%%%%%%%%%%%%%%%%%%%%%%%%%%%%%%%%%%
\section*{Sample CRediT author statement}
{\bf G. Chacon-Acosta:} Conceptualization, Methodology, Formal analysis, Writing - Original Draft, Writing - Review and Editing,Supervision.
{\bf H. Hernandez-Hernandez:} Conceptualization, Methodology, Formal analysis, Writing - Original Draft, Writing - Review and Editing,Supervision.
{\bf J. Ruvalcaba-Rascon:} Methodology, Software, Formal analysis, Resources, Writing - Original Draft,  Visualization.

\section*{Declaration of competing interests}
The authors declare that they have no known competing financial interests or personal relationships that could have appeared to influence the work reported in this paper.

\section*{Declaration of AI-Assisted Technologies}

The authors used Claude AI exclusively to improve English grammatical accuracy, clarity, and readability of the manuscript. It was not used for scientific content generation, data analysis, or interpretation of results. All scientific aspects of the work remain the sole responsibility of the authors, who have fully reviewed and approved the final manuscript.

\section*{Acknowledgments}
H.H. acknowledges support from the SECIHTI grant CBF-2023-2024-1937. We thank J. Arroyo for his comments on the manuscript.


\appendix
\section{Symplectic Structure of Quantum Variables}
\label{appendix-A}

Classical and quantum variables are symplectically orthogonal, which means that their Poisson brackets are trivial
\begin{equation}
    \bigg\{x_a, G^{a_1,b_1,\cdots,a_k,b_k} \bigg\}=\bigg\{p_a, G^{a_1,b_1,\cdots,a_k,b_k} \bigg\}=0,\label{GX}
\end{equation}
while the algebra for quantum variables is  \cite{Chacon2011}%{https://doi.org/10.48550/arxiv.1110.3337}
\begin{eqnarray}
    &\bigg\{  G^{a_1,b_1,\cdots,a_k,b_k}, G^{c_1,d_1,\cdots,c_k,d_k}\bigg\}= \nonumber \\
    &\sum\limits_{i=1}^k \biggl(a_id_i G^{a_1,b_1,\cdots,a_i-1,b_i,\cdots,a_k,b_k} G^{c_1,d_1,\cdots,c_i,d_i-1,\cdots, c_k,d_k}
    \nonumber
    \\
    &-b_i c_i G^{a_1,b_1,\cdots,a_i,b_i-1,\cdots,a_k,b_k} G^{c_1,d_1,\cdots,c_i-1,d_i,\cdots, c_k,d_k}\biggl)
    \nonumber
    \\ 
    &+\sum\limits_n\sum\limits_s\sum\limits_{e_1,\cdots,e_k} (-1)^s \left(\frac{\text{i}\hbar}{2}\right)^{n-1} \delta_{\sum_i e_i,n}
    \nonumber
    \\
    &\times\, \mathcal{K}^{n,s\{e\}}_{ \{a\},\{b\},\{c\},\{d\} } G^{a_1+c_1-e_1,b_1+d_1-e_1;\cdots; a_k+c_k-e_k,b_k+d_k-e_k},\label{GG}
\end{eqnarray}
where $n=1,\cdots,\~{N}$, and 
\begin{equation}
    \~{N}=\begin{cases}
    1, & \sum_i\left(\min[a_i,d_i]+\min[b_i,c_i] \right)\leq 1\\
    \sum_i\left(\min[a_i,d_i]+\min[b_i,c_i] \right)-1, &\sum_i\left(\min[a_i,d_i]+\min[b_i,c_i] \right)>1,
    \end{cases}
\end{equation}
 $s=0,\cdots n; 0\leq e_i\leq \min[a_i,d_i,s]+\min[b_i,c_i,n-s]$.

As for $\mathcal{K}$ it is given by
\begin{equation}
    \mathcal{K}^{n,s\{e\}}_{ \{a\},\{b\},\{c\},\{d\} }=\sum\limits_{g_1,\cdots,g_k} \frac{\delta_{\sum_ig_i,n-s}}{s!\left(n-s\right)!} \prod\limits_i \frac{\binom{a_i}{e_i-g_i}\binom{b_i}{g_i}\binom{c_i}{g_i}\binom{d_i}{e_i-g_i}}{\binom{n-s}{g_i}\binom{s}{e_i-g_i}},
\end{equation}
with $\max[e_i,s,e_i-a_i,e_i-d_i,0]\leq g_i\leq \min[b_i,c_i,n-s,e_i]$.\\

%\bibliographystyle{elsarticle-harv} 
%\bibliography{example}


\end{document}